\documentclass[a4paper,aps,amsmath,amssymb,twocolumn,longbibliography,,accepted=2024-07-14]{quantumarticle}
\pdfoutput=1
\usepackage[utf8]{inputenc}
\usepackage[english]{babel}
\usepackage[T1]{fontenc}
\usepackage{amsmath}
\usepackage{hyperref}

\usepackage{tikz}
\usepackage{lipsum}

\usepackage[utf8]{inputenc}

\usepackage{tikz}

\usetikzlibrary{decorations.markings}

\usepackage{amsmath}  
\usepackage{url}
\usepackage{hyperref}
\usepackage{mathtools}
\usepackage{siunitx}
\usepackage{multirow}
\usepackage{microtype}
\usepackage{array}
\usepackage{tabulary}
\usepackage{longtable}
\usepackage[version=4]{mhchem}
\usepackage{graphicx}
\newcolumntype{K}[1]{>{\centering\arraybackslash}p{#1}}

\usepackage{graphicx}   
\usepackage{verbatim}   
\usepackage{color}      
\usepackage{subfigure}  
\usepackage{hyperref}   
\newcommand{\btxt}[1]{{\color{black} #1}}

\usepackage{lipsum}

\begin{document}

\title{Deep learning of many-body observables and quantum information scrambling}

\author {Naeimeh Mohseni}
\affiliation{Physics Department, Friedrich-Alexander Universität of Erlangen-Nuremberg, Staudtstr. 7, 91058 Erlangen, Germany}
\affiliation {Max-Planck-Institut f{\"u}r die Physik des Lichts, Staudtstrasse 2, 91058 Erlangen, Germany}

 \author{Junheng Shi}
\affiliation{New York University Shanghai, 1555 Century Ave, Pudong, Shanghai 200122, China}

\author{Tim Byrnes}
\affiliation{New York University Shanghai, 1555 Century Ave, Pudong, Shanghai 200122, China}

\affiliation{National Institute of Informatics, 2-1-2 Hitotsubashi, Chiyoda-ku, Tokyo 101-8430, Japan}
\affiliation{Department of Physics, New York University, New York, NY 10003, USA}
\author{Michael J. Hartmann}
\affiliation{Physics Department, Friedrich-Alexander Universität of Erlangen-Nuremberg, Staudtstr. 7, 91058 Erlangen, Germany}
\affiliation {Max-Planck-Institut f{\"u}r die Physik des Lichts, Staudtstrasse 2, 91058 Erlangen, Germany}

\begin{abstract}
Machine learning has shown significant breakthroughs in quantum science, where in particular deep neural networks exhibit remarkable power in modeling quantum many-body systems.
Here, we explore how the capacity of data-driven deep neural networks in learning the dynamics of physical observables is correlated with the scrambling of quantum information. We train a neural network to find a mapping from the parameters of a model to the evolution of observables in random quantum circuits for various regimes of quantum scrambling and test its \textit{generalization} and \textit{extrapolation} capabilities in applying it to unseen circuits. Our results show that a specific type of recurrent neural network can generalize its predictions within the system size and time window that it has been trained on across both, localized and scrambled regimes.  Moreover, the considered neural network succeeds in \textit{extrapolating} its predictions beyond the time window and system size that it has been trained on for models that show localization, but not in scrambled regimes.

\end{abstract}
\maketitle
\section{Introduction\label{sec:introduction}}

Non-equilibrium dynamics of quantum many-body systems \cite{eisert2015quantum, PhysRevA.103.023713} plays an essential role in many fields across physics, ranging from ultra-cold atoms \cite{gross2017quantum, bernien2017probing} to strongly correlated electron materials \cite{morosan2012strongly}, quantum information processing \cite{anderlini2007controlled}, and quantum computing \cite{boixo2018characterizing}.  Due to the exponential scaling of the Hilbert space dimension, a complete description of a generic many-body state requires an exponential amount of classical resources and thus becomes intractable already at moderate system sizes.
The nature of entanglement and correlations together with the way they spread throughout the system are the main source for this computational complexity.
Hence, substantial research is being conducted to understand the representational power of classical methods and its relation to entanglement growth 
\cite{jia2020entanglement, gao2017efficient, PhysRevX.7.021021, cirac2021matrix, hamza2012dynamical}.

Classical machine learning algorithms have exhibited an impressive ability to find high-accuracy approximations for desired quantities of quantum many-body systems, especially for problems that do not permit numerically exact solutions \cite{huang2022learning, PhysRevLett.125.100503, PhysRevLett.122.250502, dawid2022modern}. In particular, the challenging task of computing real-time-evolutions of many-body dynamics has been addressed using both data-driven learning methods \cite{banchi2018modelling, herrera2021convolutional,nmohseni2021deep, mohseni2022deep, schmidt2024transfer} and direct calculation methods. Especially for the latter, the neural network wave function ansatz \cite{carleo2017solving, schmitt2018quantum, PhysRevLett.122.250502,  lange2024architectures, medvidovic2024neural}, where neural networks find an efficient representation for the wave function, has entailed large interest. 
 \btxt{In a few recent works, the representational power of the neural network wave function ansatz and its connection to the entanglement features of corresponding quantum states has been explored \cite{jia2020entanglement, gao2017efficient, PhysRevX.7.021021}. However, the role of entanglement in the performance of classical data-driven neural networks remains relatively unexplored. Since data-driven methods are distinct from neural network wave functions, the significance of entanglement in their representational power may also differ significantly.}

Understanding the connection between the power of data-driven methods in learning the dynamics of physical observables and the scrambling of quantum information in these systems is very important since these methods eliminate the need for expensive direct calculations and can thus form a powerful classical tool to predict the dynamics of observables in quantum many-body systems.

Here we explore this connection in an investigation of the dynamics generated by random quantum circuits, which allow us to interpolate between various regimes of quantum scrambling.  We train a neural network to predict directly the time evolution of physical observables for given time traces of control fields and parameters of the model. In this approach, the neural network finds an efficient representation of the model just by monitoring the data (e.g. expectation values of observables for various evolution times)  without having information about the underlying physics or utilizing any explicit assumptions about the considered model.  We observe that the neural network we use succeeds in  \textit{generalizing}  its predictions, within the system size and time window that it has been trained on in both, localized and scrambled regimes. In contrast, for \textit{extrapolating}  its prediction beyond the time window and system size that it has been trained on, it only succeeds for the many-body localized models. 

The paper is organized as follows. 
We first explain the physical model, i.e. the quantum circuits, we consider (section II) to confirm and discuss that it indeed exhibits regimes of localization and information scrambling. Section III then explains the learning strategy that we apply for training the neural network before we present our results for the generalization and extrapolation of the network predictions in section IV. Finally, we present our conclusions and an outlook.

\section{Physical Model \label{regimes}}

\begin{figure*}[ht!]
\centering
\includegraphics[width=1\linewidth]{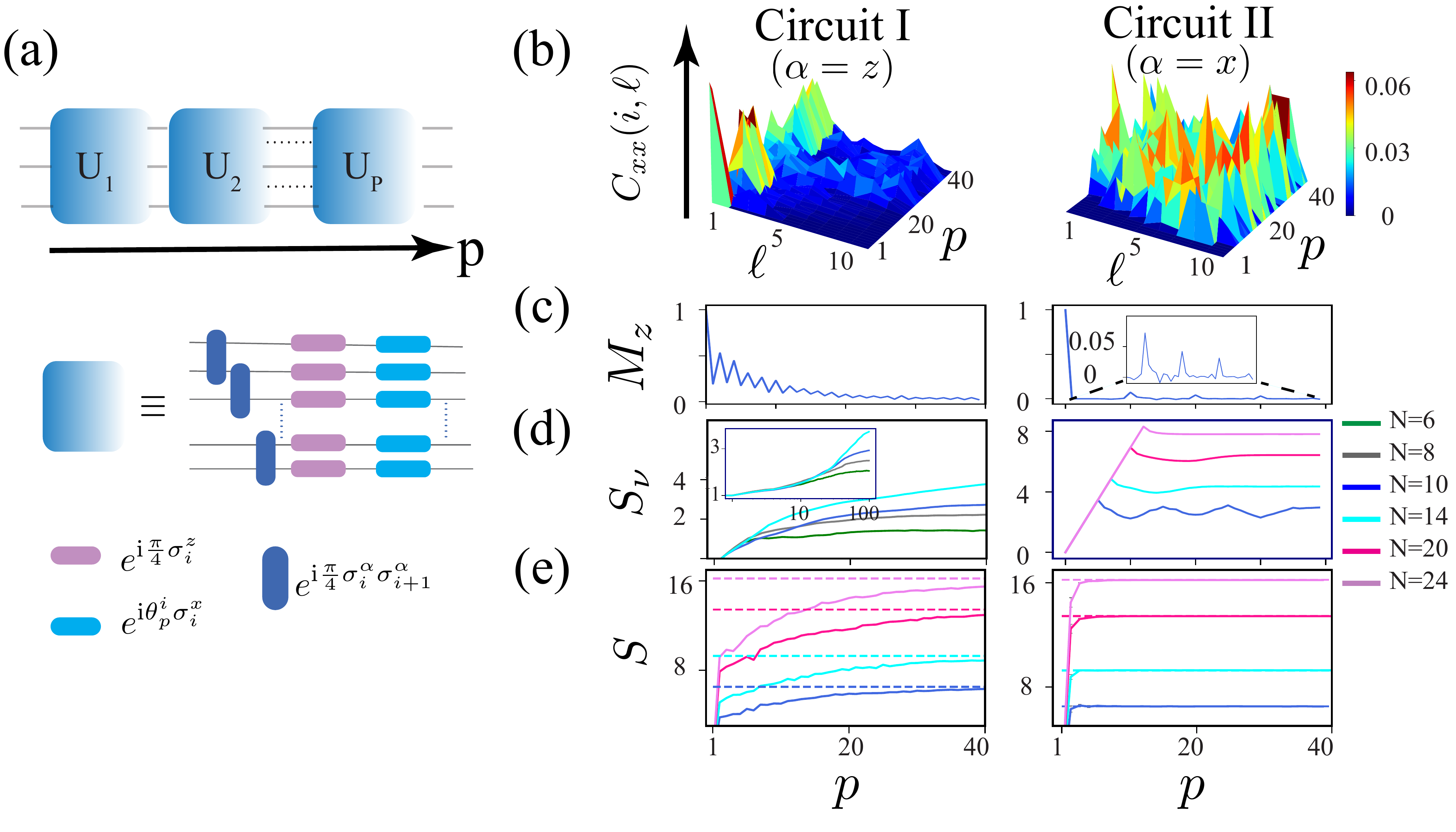} 
\caption {Physical model: (a) Schematic representation of the random circuit.    The circuit is made of modules shown as blue cells. Each module is made of three layers shown at the bottom. 
  (b)  Two-point correlators  $C_{xx}(i,\ell)=|\langle \sigma^x_{i+\ell} \sigma_i^x \rangle- \langle \sigma^x_{i+\ell}\rangle \langle \sigma_i^x \rangle  |^2$ versus circuit depth and distance $\ell$ for the two different choices of two-body gates, circuit I (left) and circuit II (right), for $N=10$. (c)-(e) Magnetization, von-Neumann entropy, and entropy versus circuit depth for different system sizes for both, circuit I (left column) and circuit II (right column). Each curve represents an average over 20 circuit realizations.  Dashed lines in panel (e) present the entropy of the ideal Porter-Thomas distribution. All shown results are for realizations of the circuits that are homogeneous in space, where $C_{xx}(i,\ell)$ is the same for all qubits $i$.
 \label{fig2}}
\end{figure*} 

As we are interested in exploring the correlation between the capacity of data-driven deep neural networks in learning the dynamics and the scrambling of quantum information, we design our 1D random circuit such that it can produce dynamics in two distinct regimes: one regime for which quantum information localizes and another regime where quantum information is scrambled. Scrambling for closed quantum systems describes a process for which initially localized quantum information spreads out throughout the system and randomizes the quantum state such that it makes the quantum information inaccessible to local observables  \cite{PhysRevLett.120.130502}. In contrast, for many-body localized systems, information about the initial state  can be extracted from a subsystem.

In Fig. \ref{fig2}a, we show the schematic representation of our circuit which is made of P modules, shown as blue cells, described with unitary operators $U_p$ with $p=1,2 \dots, \text{P}$,
\begin{align} \label{eq:circuit}
    U = & \prod_{p=1}^{\text{P}} U_p, \: \: \text{where} \: \:
    U_p = \prod_{i=1}^N e^{-i \frac{\pi}{4}\sigma_i^\alpha \sigma_{i+1}^\alpha} e^{-i \frac{\pi}{4} \sigma_i^z} e^{-i \theta^i_p \sigma_i^x}
\end{align} 
where the index $i$ labels the qubits and we consider closed boundary conditions. $N$ denotes the number of qubits. Each cell is made of three layers. The first layer  is made of two-body gates $e^{-i \frac{\pi}{4}\sigma_i^\alpha \sigma_{i+1}^\alpha}$, for which we consider the two cases $\alpha=z$, that we call circuit I,  and  $\alpha=x$,  that we call circuit II. The second and third layers are formed by single-qubit gates $e^{-i \theta_p^i \sigma_i^x}$ and $e^{-i \frac{\pi}{4} \sigma_i^z}$. The rotation angles $\theta ^i_p \in [0,\pi]$ are our input parameters, which are chosen at random and can thus introduce disorder. We consider cases where the $\theta^i_p$ are inhomogeneous in both, space and time, and where they are just inhomogeneous in time but homogeneous in space ($\theta^i_p = \theta^j_p$). To generate the random trajectories for the $\theta$s, we use a random Gaussian process \cite{liu2019advances}, see supplemental material information Sec. I for more details.   

Circuit I with  $\alpha=z$ creates many-body localized (MBL) dynamics, while circuit II with  $\alpha=x$ creates thermalizing dynamics, where information scrambling happens. 
MBL and thermalized systems have unique characteristics that distinguish them. Here we check a few of these, for both choices of two-body gates, circuit I and circuit II, to confirm that the dynamics of our circuit is scrambled or localized.

MBL phases are characterized by an exponential decay of two-body correlations 
\cite{imbrie2016many} while such correlators do not decay when the system thermalizes. Localized dynamics is also characterized by a slow, power-law relaxation of local (e.g. single qubit) observables towards stationary values that are highly dependent on the initial condition \cite{PhysRevB.90.174302}. In contrast, local observables decay exponentially towards stationary values with only weak dependence on initial conditions where information scrambling occurs.  Moreover, MBL systems are characterized by slow logarithmic growth of entanglement entropy starting from a low entanglement or product state and they saturate to a value that obeys a volume law.  
In contrast, when the system thermalizes, the entanglement entropy grows linearly and saturates to a value that is system-dependent and obeys a volume law.  

To monitor how correlations build up in our circuits, we investigate the evolution of two-point correlators  
$C_{\gamma \beta}(i,l)=|\langle \sigma^\gamma_{i+\ell} \sigma_i^\beta \rangle- \langle \sigma^\gamma_{i+\ell}\rangle \langle \sigma_i^\beta \rangle  |^2$ where the expectation values are taken over the wave function at each circuit depth, $\gamma ,\beta = x,y,z$, and we chose the input parameters homogeneous in space, $\theta_p^i = \theta_p^j$ . For circuit I, the evolution of  $C_{xx}$ exhibits localization in space indicating that the wave function becomes localized in some region of space and decays exponentially far away from that region, see Fig. \ref{fig2} (b). This localization persists almost for the entire shown circuit depths. On the other hand, for circuit II, long-range correlations build up already after very short circuit depth. 
As for local observables, we look at the magnetization calculated as $M_z=\frac{1}{N}\sum_{i=1}^N\sigma_i^z$ where $N$ denotes the number of qubits.  Fig. \ref{fig2} (c) shows the average of magnetization over 20 realizations for both types of circuits.  The magnetization collapses polynomially with the circuit depth for circuit I while it decays exponentially for circuit II.

 To study the growth of entanglement, we calculate the von Neumann entropy of the reduced density matrix $\rho_r$ for half of the circuit defined as $S_v=- \text{Tr} [\rho_r \ln \rho_r]$.  We also calculate   entropy  defined as  $S=- \sum_{i=1}^{N}P_i \ln P_i$ where $P_i = |\langle \psi|i\rangle|^2$ represents the probability of finding the state $|\psi\rangle$ of the system in the $i$-th  computational basis state $|i\rangle$. We compare for each regime the entropy of our circuit with the entropy of a perfect  Porter-Thomas distribution which equals $M\ln(2)-1 -\gamma$ with $\gamma$ representing the Euler’s constant \cite{boixo2018characterizing}.  The Porter-Thomas (PT) distribution is characteristic of chaotic dynamics for which the fractional of the configurations that have probabilities in a given range $[p,p+dp]$ decays exponentially as $p 2^{2 N} e^{-2^N p} dp$   and it is unlikely to simplify a circuit substantially when its probability distribution converges to PT \cite{boixo2018characterizing}.  An entropy $S$ converging to the entropy of PT distribution implies that thermalization occurs and dynamics become chaotic.

For circuit I, the von-Neumann entropy  (Fig. \ref{fig2} (d)) starts with rapid linear growth for a quite small circuit depth and then is followed by slow logarithmic growth before it eventually saturates. The saturation value $\kappa L$ appears to obey a volume law with $\kappa$ smaller than its maximum value of $\ln 2$, where $L=N/2$ is the length of the partition.  The inset shows the growth of von-Neumann entropy for a larger circuit depth (semi-log scale) where saturation for the shown system sizes can be seen clearly.  The duration of logarithmic growth increases with system size.  Here we look at the dynamics before saturation occurs.  For circuit II, the von-Neumann entropy shows a fast linear growth which then rapidly saturates to $\kappa L$. The linear growth of the von Neumann entropy reflects the spreading of correlations at a finite speed before saturating because of the finite size of the system. The saturation value follows a volume law with $\kappa$ being close to its maximum value of $\ln 2$ which is a signature of thermalization and chaos meaning that all degrees of freedom become highly entangled with each other throughout the quantum evolution.

For circuit I, it is also evident that, the larger the system size gets, the deviation of the probability distribution of the circuit from the PT distribution at large circuit depths becomes more evident, see Fig. \ref{fig2} (e). In contrast, for circuit II,  the entropy converges to the entropy of the perfect PT distribution quite fast after a few modules, see Fig. \ref{fig2} (e).

\section{Learning strategy }
We now explore the learning capacity of  a data-driven learning approach in which a neural network learns to predict the physical observables directly,  rather than learning the wave function. Our choice is motivated by the fact that finding an efficient representation for the quantum state is computationally expensive, while for many goals, we do not need the full wave function but only the expectation values of a selected subset of observables.  Moreover, the existence of an efficient representation of the quantum state does not imply that physical observables can be calculated efficiently, since the latter may involve complex index contractions \cite{gao2017efficient}. Our direct training on physical observables forgoes such needs to deal with the exponentially large state vector itself.

\begin{figure*}[t]
\centering
\includegraphics[width=0.9\linewidth]{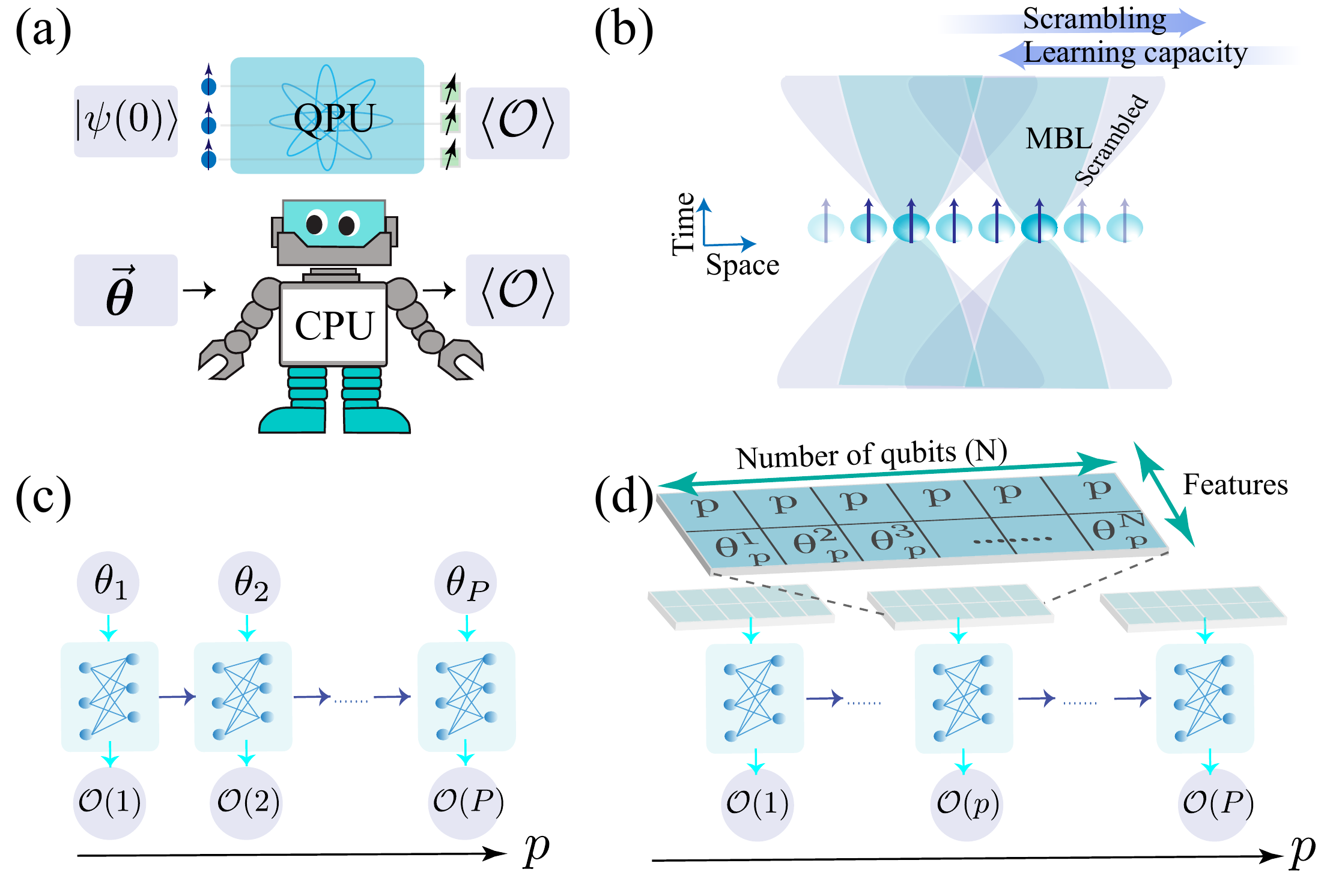} 
\caption {Schematic representation of the applied learning strategy. (a) The neural network learns a simple and efficient representation for the equations of motion when provided with the parameters of the model (shown with $\boldsymbol{\vec{\theta}}$) just by observing a subset of physical observables but without knowing the unitary operator that describes the model. (b) The propagation of information in a many-body system forms a light cone. For MBL models the radius of the light cone grows logarithmically while it propagates polynomially for scrambled models. Two qubits do not affect each other if one lies outside
of the light cone of the other. The representation learned by the neural network is reliable beyond the system size and time window that it has been trained on where information scrambles slowly (logarithmically) and the dynamics is localized.  (c) Schematic representation of the LSTM neural network. The input of the network at each circuit depth $p$ is a set of random parameters $\theta_p$  as well as the circuit depth $p$. At each circuit depth $p$ the network provides as output the dynamics of desired observables $\langle \mathcal{O}(p) \rangle$. \btxt{(d) Schematic representation of the 1D CONVLSTM neural network. The input of the network at each circuit depth $p$ is shown in the green rectangle which includes the parameter $\theta^i_p$ for different qubits as well as the circuit depth $p$. At each circuit depth $p$ the network provides as output the dynamics of the desired observables $\langle \mathcal{O}(p) \rangle$}.  The horizontal blue arrows in (c) and (d) indicate the content of the internal neural memory being passed to the next time
step.
 \label{fig1}}
\end{figure*}

In general, learning the dynamics from partial observations without having access to a full representation of the wave function is a non-trivial task.  The reason is that, for a generic many-body model, the evolution of each observable depends on the evolution of many or even all other observables, as becomes evident from the Heisenberg picture equations of motion. From this point of view, one would expect that predicting the dynamics of one observable can require knowledge of the full wave function.  In contrast, the neural network approach we use aims at finding an effective representation of the equations of motion just by observing a subset of observables (Fig. \ref{fig1} (a)).  The first question that we are interested in answering here is whether a neural network succeeds in finding such an effective representation for models with different levels of complexity. By complexity, we mean the way that information is scrambled.  The next interesting question is whether the representation found by the neural network for a given system size and time window can even be used to predict the dynamics for larger system sizes and longer times than the network has been trained on despite the typical generation of entanglement between increasingly distant regions as time progresses.  We observe that such extrapolations are only successful 
when information scrambling  
 occurs slowly, which is the characteristic of the many-body localized models.

\textbf{Neural network architecture:} We apply a particular type of recurrent neural network called a long-short-term memory (LSTM) neural network for this task. Our choice is motivated by the fact that this architecture naturally respects the fundamental principle of causality, which makes them well-suited to represent differential equations (equations of motion). Moreover, this architecture is known for capturing both
long-term and short-term dependencies which gives it the power to handle complex non-Markovian dynamics. Importantly, it also permits extrapolation in time as it can be used for varying input sizes.  To explore the possibility of extrapolating the dynamics of the observables to larger system sizes,  we combine our LSTM network with a convolutional neural network so-called convolutional long-short-term memory (CONVLSTM) neural network \cite{shi2015convolutional}.

\textbf{Training:} In Fig. \ref{fig1} (c) and (d),  we represent the schematic of our LSTM and CONVLSTM  networks, respectively.  We feed as input $p$ and the parameters $\theta_p^i$, which determine the gates applied to the qubits see Eq. (\ref{eq:circuit}). In both cases, the neural network provides as output the desired observables for the considered circuit depth.  See Ref. \cite{nmohseni2021deep}  for more details about LSTM and COVLSTM architectures and how they decide the flow of information in and out at each step. 

We always start from a product state where all qubits are prepared in the +1 eigenstate of the $\sigma_z$ operator. As an example, we here train the network on first and second-order moments of spin operators  ($\langle \sigma^\alpha_{i} \rangle, \langle \sigma^\alpha_{i} \sigma_{i+\ell}^\beta \rangle$)  with $\alpha,\beta=x,y,z$ as many interesting physical observables can be obtained from these quantities. Also, these observables can be measured in experiments meaning that one can even train the neural network on data obtained from experiments.
The cost function that we use to train our neural network is defined as
\begin{equation} \label{eq:mse}
{\textrm{MSE}}= \overline{|\langle \mathcal{O} \rangle_{\rm NN} (p) - \langle\mathcal{O}\rangle_{\rm true} (p)|^2 }
\end{equation}
where the bar shows the average over all samples and circuit depths. $\langle \mathcal{O}(p) \rangle$ denotes the expectation value of the desired observables at circuit depth $p$. Note that for the case where we combine our LSTM network with CNN, we feed our input with a spatio-temporal structure to the network.

Our approach differs from works that apply recurrent or convolutional neural networks to learn the wave function  \cite{banchi2018modelling, herrera2021convolutional} as our neural network directly learns the dynamics of physical observables and therefore can also be applied to large system sizes, for which storing an entire wavefunction requires exceeding amount of memory. There are some other works that also use neural networks to predict the dynamics of physical observables. But these consider only a single qubit \cite{flurin2020using} or aim to learn the dynamics of a single qubit by considering all other qubits as a quantum environment  \cite{mazza2021machine}. In contrast, our network learns the dynamics of all qubits simultaneously. Another difference is that in most of these works, the neural network learns to predict the dynamics for longer times by having the short-time evolution of a system as input \cite{banchi2018modelling, mazza2021machine}, and that mostly works fine where parameters of the model do not change with time.    In contrast, in our strategy, the neural network finds a mapping from the parameters ($\theta_p^i$) of the model, that are always inhomogeneous in time, to the dynamics of physical observables. More important than that there is no systematic study to discuss how the learning capacity of a data-driven method in learning many-body dynamics is connected with the scrambling of quantum information and where are the regimes that the representation found by the neural network is still reliable beyond the system size and the time-window that it has been trained on.


\section{Results}
In this section, we discuss the performance of the neural network in learning the many-body dynamics for the two circuits introduced in Sec. \ref{regimes}. We first evaluate the performance of the network on unseen realizations of the circuit for the circuit depth and system size that it has been trained on to evaluate its generalization power. Then we explore the power of our neural network in extrapolating its prediction to system sizes and circuit depths that it has not been trained on.

\begin{figure}[t!]
\centering
\includegraphics[width=1.03\linewidth]{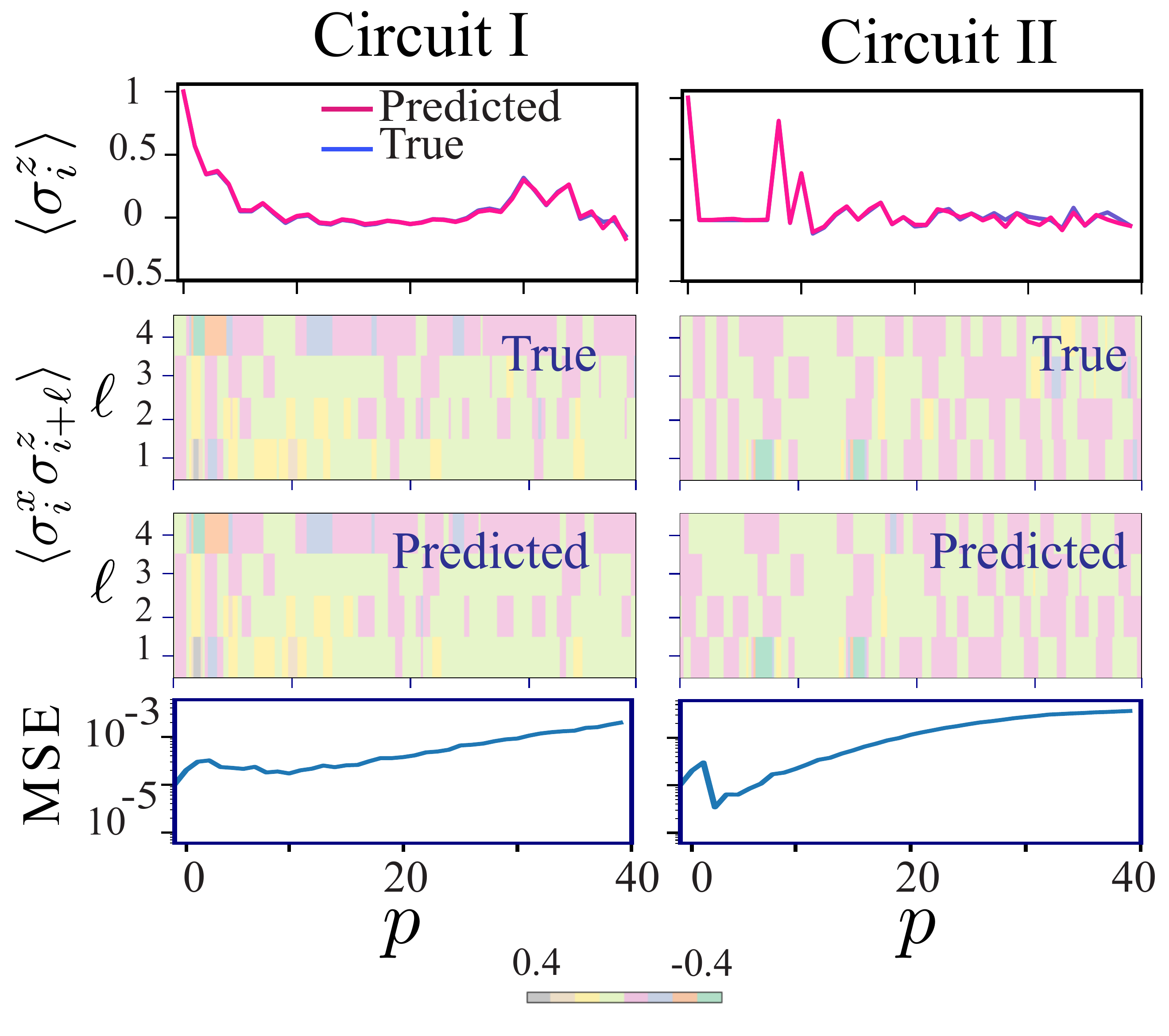} 
\caption {Generalization power of the LSTM network: Two separate LSTM neural networks are trained independently for circuits I and II, using random realizations of each circuit with $N=8$ qubits and $p\in[1,40]$. Subsequently, they are evaluated independently to predict the dynamics of observables on unseen realizations of each circuit with $N=8$ qubits and $p\in[1,40]$.   The performance of these 
 neural networks in predicting  $\langle \sigma^z_i  \rangle $ and $\langle \sigma^x_i \sigma^z_{i+\ell} \rangle$  for a typical realization of both circuits is shown. The lower panels show the MSE (c.f. Eq. (\ref{eq:mse})), averaged over 1000 realizations of the circuit. For each realization of the circuit, the neural network is trained on 30 observables \btxt{ (3 first-order moments and 27 second-order moments of spin operators) simultaneously. Both circuits are chosen to be homogeneous in space ($\theta^i_p = \theta^j_p$), hence $\langle \sigma_i^z \rangle$ and $\langle \sigma_i^x \sigma_{i+\ell}^z \rangle$ are equal for all qubits $i$.}
 \label{fig3}}
\end{figure}

\textbf{Generalization:}
 We train and evaluate our neural network on a system of size $N=8$ for random realizations of each circuit separately, where $p\in[1,40]$.  For both circuits, the parameters $\theta_p^i$ are chosen inhomogeneous in time but homogeneous in space ( $\theta_p^i = \theta_p^j$)\btxt{,  hence $\langle \sigma_i^{\alpha}\rangle$ and $\langle \sigma_i^{\alpha} \sigma_{i+\ell}^{\beta}\rangle$ where  $\alpha,\beta=x,y,z$ are equal for all qubits $i$.  The neural network is trained simultaneously on the dynamics of 30 observables  (3 first-order moments and 27 second-order moments of spin operators)}. See the supplemental material for more information about the training set size and the neural network structure.
 In Fig. \ref{fig3}, we show the predicted and true dynamics of  $\langle \sigma^z_i  \rangle $ and $\langle \sigma^x_i \sigma^z_{i+\ell} \rangle$ for one typical realization of the circuit.  As can be seen, the network is able to learn the dynamics of these observables with high precision for both implementations of the circuit. Yet the precision of predictions at larger circuit depths is higher for circuit I in comparison to circuit II. The lower panels show the MSE, defined in Eq. (\ref{eq:mse}), where we average over 1000 realizations of each circuit.

\btxt{In the context of these results, one should note that the capability of classical learning methods in sampling from random quantum circuits in different regimes has been explored demonstrating that classical learning tools fail in sampling in the regime where the probability distribution converges to a PT distribution and quantum information is scrambled \cite{niu2020learnability, hinsche2021learnability, rudolph2022synergy}. It is thus interesting to see that our learning strategy using a recurrent neural network succeeds in learning the dynamics of desired physical observables in this regime. This is particularly relevant as learning observables can be even more useful.
}

\begin{figure}[t]
\centering
\includegraphics[width=0.7\linewidth]{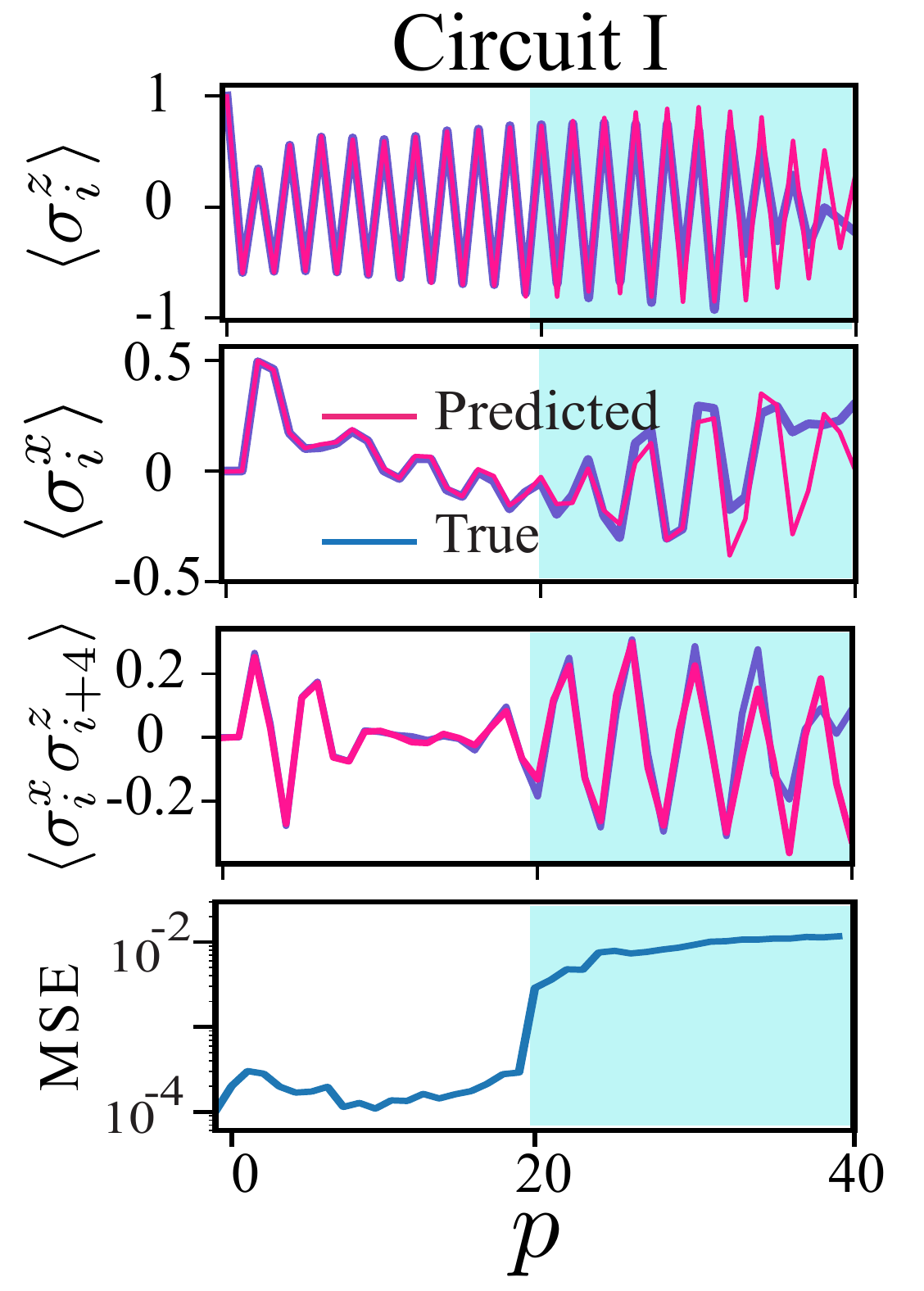} 
\caption {Extrapolation power of the LSTM neural network in circuit depth for circuit I. The LSTM neural network is trained on the physical observables for random realizations of the circuit I on system size $N=8$ with $p \in[1,20]$. It is then evaluated on unseen realizations of the circuit  with $N=8$ and $p \in[1,40]$. The performance of the network in generalization ($p \in[1,20]$), as well as extrapolation in circuit depth (highlighted with blue, $p \in[20,40]$) for a typical realization of the circuit, is shown. The lower right panel shows the MSE (c.f. Eq. (\ref{eq:mse})) averaged over 1000 realizations of the circuit. For each realization of the circuit, the neural network is trained on 30 observables simultaneously \btxt{ (3 first-order moments and 27 second-order moments of spin operators). The circuit is chosen to be homogeneous in space ($\theta^i_p = \theta^j_p$), hence $\langle \sigma_i^z \rangle$, $\langle \sigma_i^x \rangle$ and $\langle \sigma_i^x \sigma_{i+4}^z \rangle$ are equal for all qubits $i$}.
 \label{fig4}}
\end{figure}

\textbf{Extrapolation in circuit depth:} We also investigate the power of our LSTM neural network in extrapolating the dynamics of monitored physical observables to larger circuit depths than it has been trained on. Here we observe that the trained neural network succeeds in extrapolation just for circuit I where MBL occurs.  We train the neural network simultaneously on the dynamics of observables for $p \in [1,20]$ and evaluate it on unseen realizations $\theta_p^i$ of the circuit with $p \in [1,40]$. In Fig. \ref{fig4}, the blue highlighted regions present circuit depths that the neural network has not been trained on and thus extrapolates to. We interpret the observed behavior as follows.

Even though the dynamics is unitary and invertible, the information about the initial state becomes, in scenarios where information scrambling occurs, inaccessible to local observables 
and recovering that information would require measuring global operators \cite{PhysRevLett.120.130502}. Therefore, the neural network fails here in extrapolating the dynamics of local observables as it loses locally information about the past. In contrast, in regimes where MBL happens, the information encoded in the initial state is retained in local observables which therefore can govern the dynamics at longer times. In such models,  an extensive set of local integrals of motion describes the dynamics. Therefore, success in extrapolation may suggest that the neural network learns such local integrals of motion just by observing a subset of local observables. This can explain why the neural network succeeds in predicting the dynamics for larger circuit depths than it has ever been trained on despite the typical generation of entanglement between increasingly distant regions as time progresses.  It is computationally hard to further inspect this conjecture, that the neural network may learn the local integrals of motion. The reason is that calculating the local integrals of motion for our model is very complicated. Also, it is very challenging to inspect what exactly the neural network learns. 

\begin{figure} [t]
\centering
\includegraphics[width=0.75\linewidth]{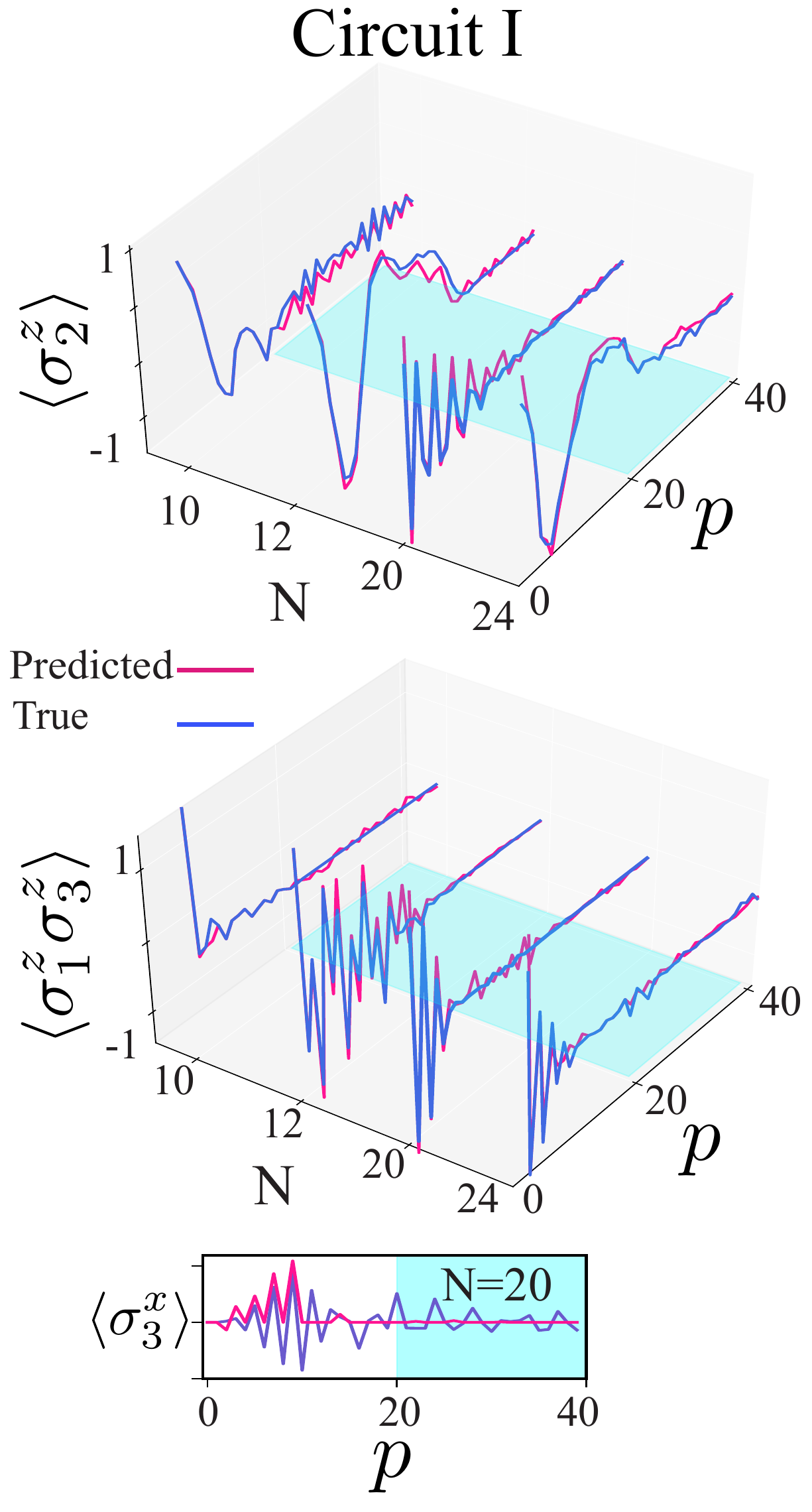} 
\caption {Extrapolation power of CONVLSTM in system size and circuit depth for circuit I. The network is trained on system size $N=8$ and $p \in[1,20]$ on the dynamics of desired observables and evaluated on unseen samples with $N=10,12,20,24$ and $p \in[1,40]$. The performance of the network in predicting a few observables is shown. The neural network is trained on inhomogeneous (both in time and space)  realizations of the circuit I ($\theta^i_p \neq \theta^j_p$) \btxt{  on 12 first-order moments and 27 second-order moments of spin operators.  }
 \label{fig7}}
\end{figure}

\begin{figure}[t]
\centering
\includegraphics[width=1\linewidth]{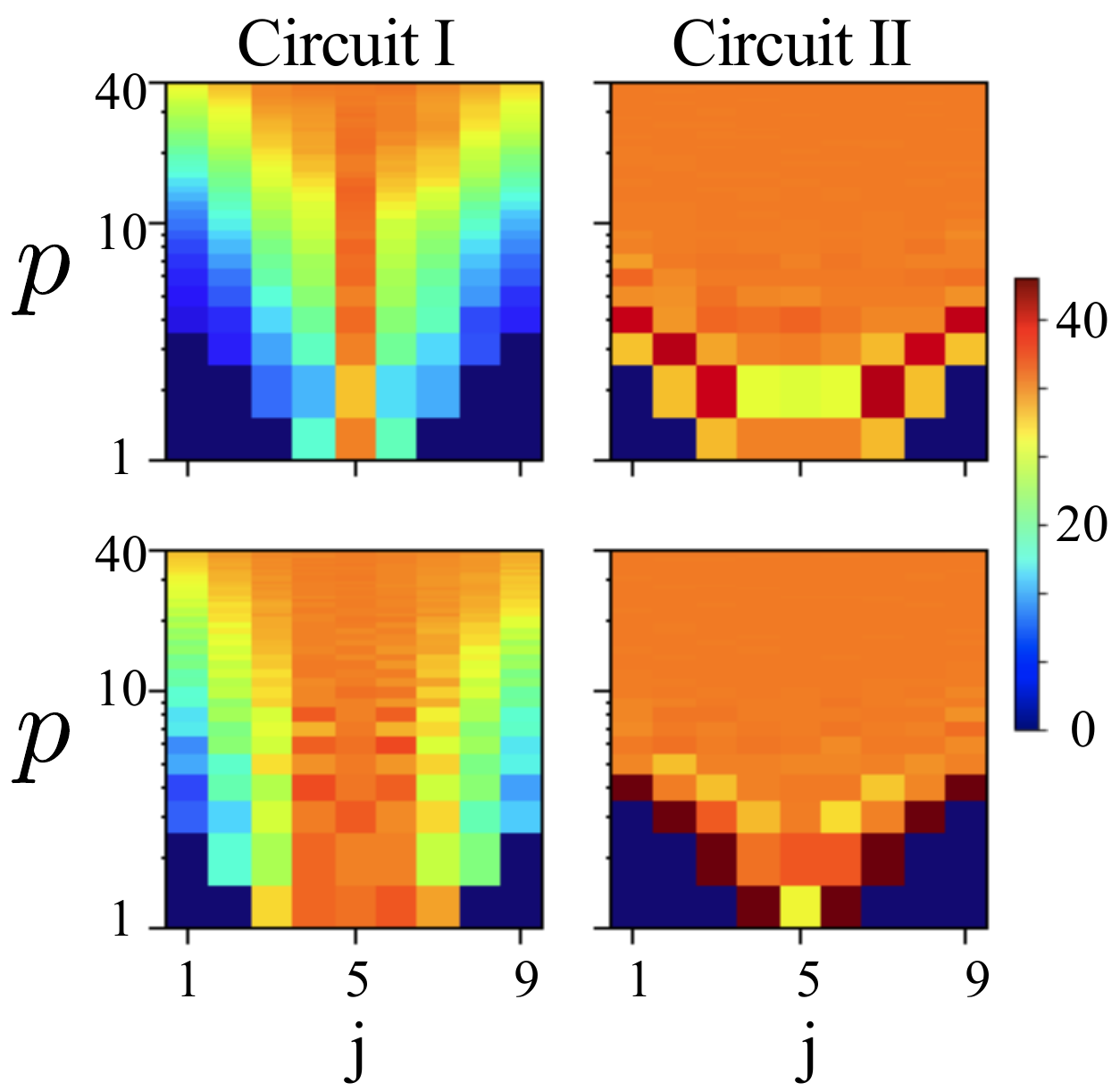} 
\caption {Light-cone spreading of quantum information. The top panels show  $ \left\|\left[\sigma_5^z, \sigma_j^z(p)\right]\right\|_F$ and the lower panels show $\left\|\left[\sigma_5^x, \sigma_j^x(p)\right]\right\|_F$  for $N=9$ averaged over 40 realizations for each panel.  Both circuits are inhomogeneous in time and space ($\theta^i_p \neq \theta^j_p$). 
 \label{fig66}}
\end{figure}

\textbf{Extrapolation in system size:} For exploring the possibility of extrapolating the predictions of the neural network to system sizes beyond those that it has been trained on,  we choose our circuit to be inhomogeneous both in time and space ($\theta^i_p \neq \theta^j_p$). We also combine our LSTM neural network with a 1D CNN  network \cite{xingjian2015convolutional}. This architecture is designed  for data with spatio-temporal structure \cite{shi2015convolutional}, where the CNN is applied to deal with the spatial structure of the input and the LSTM  keeps track of the evolution. See Supplemental Material of Ref. \cite{nmohseni2021deep} for more technical details about this architecture.  

Obviously, the dynamics of a given qubit is affected by increasingly many other qubits as time progresses. One might thus expect that it should be challenging for a neural network to find some effective description that can include the influences of more qubits than it has been trained on. We observe that the neural network succeeds in generalizing and  extrapolating the dynamics to larger system sizes for circuit I where MBL occurs while it fails for circuit II where scrambling occurs. However, even  for circuit I, the precision of the neural network in learning local observables that contain $\sigma^x_i$  is generally lower than other observables, and the neural network can only learn their dynamics  for smaller circuit depths. This can be clearly seen in Fig. \ref{fig7}  where  a CONVLSTM  is trained on system size $N=8$ with $p \in [1,20]$ and is evaluated on $N=10,12,20,24$ with $p \in [1,40]$ for a few typical realizations of the circuit I.

We interpret these observations as follows. For observables, for which the neural network can extrapolate the dynamics, the support of their operators in a Heisenberg picture representation remains well localized. Therefore, qubits that are far apart (in comparison to the localization length) do not contribute significantly to the dynamics of these local observables. In this case, increasing the system size does not affect local observables notably even though the entanglement entropy may still grow. 
To confirm our interpretation, we calculate the out-of-time-order correlator (OTOC) for an operator $\mathcal{O} $  defined as $\left\|\left[\mathcal{O}_0(0), \mathcal{O}_j(p)\right]\right\|_F$ where $\mathcal{O}_j(p)=U \mathcal{O}_j (0) U^{\dagger}$ and $\left\|.\right\|_F$ represents the Frobenius norm.  This OTOC is often used to characterize the information scrambling and chaos. For a localized model, the propagation of information forms a light cone where the OTOC  is non-negligible inside this light cone whose radius is proportional to $\log(t)$ and decays exponentially with distance outside the light cone. In contrast for the cases where scrambling occurs the OTOC shows a power-law light cone  \cite{nahum2018operator}.

In Fig. \ref{fig66}, we show $ \left\|\left[\sigma_5^{\alpha}, \sigma_j^{\alpha}(p)\right]\right\|_F$  for $\alpha=x,z$ and $N=9$ with $j=1,2,3,...,9$.  As can be seen for circuit I,  $\sigma^x_i$ spreads faster after a short circuit depth which explains why the neural network learns the dynamics of $\sigma^x_i$ observables with lower precision and for smaller circuit depth in comparison with other observables such as $\sigma^z_i$ which remains well localized.  In the right column, we also show the same for circuit II where scrambling happens. It is obvious that after a short circuit depth, both observables spread fast.

 \section{Conclusion and outlook\label{conclusion}}
In this work, we show that data-driven recurrent neural networks succeed in learning the dynamics of many-body systems--- within the trained time window and system size---  in both MBL and scrambled regimes. Learning the dynamics of physical observables for scrambled dynamics is of special interest as classical learning tools are known to fail in sampling from the output of quantum circuits in this regime. Our results show that while neural networks fail to learn the full information about the wave function they can still learn the dynamics of desired physical observables, a capability that is even more valuable than predicting the wave function in many applications.  We also observe that a trained convolutional recurrent neural network succeeds in extrapolating the predictions beyond the trained time window and system size for cases where MBL occurs while it fails in regimes where information scrambling occurs. We attribute this observation to the fact that for MBL models the dynamics is governed by local integrals of motion which do not change in time and have a localized support in a Heisenberg picture representation so that distant qubits do not contribute to local observables’ dynamics.

\btxt{
Further explorations of the potential of data-driven methods in learning many body dynamics could incorporate transformers \cite{vaswani2017attention, zhang2023transformer} or integrate recurrent neural networks with transformers. Transformers, with their attention mechanisms, may be useful in addressing the challenges posed by scrambled dynamics. This mechanism could empower transformers to more effectively capture long-range dependencies, potentially surpassing LSTMs in such scenarios. Nevertheless, empirical evidence for the success of this idea would be needed via thorough investigation, especially within the framework of data-driven methods.}

 In this work, we trained our neural network on the data generated from numerical simulations. An interesting perspective for future work would thus be to train the neural network on the data generated by actual experiments. We briefly comment on the resources required for such an investigation in the supplemental material Sec. II. 

\begin{acknowledgments}
 N.M  thanks Xiangyi Meng,  Hongzheng Zhao, and Sharareh Sayyad for valuable discussions. This is part of the Munich Quantum Valley, which is supported by the Bavarian state government with funds from the Hightech Agenda Bayern Plus. It also received funding from the European Union's Horizon 2020 research and innovation program under Grant Agreement No. 828826 ``Quromorphic.''
T. B.  is supported by the National Natural Science Foundation of China (62071301); NYU-ECNU Institute of Physics at NYU Shanghai; the Joint Physics Research Institute Challenge Grant; the Science and Technology Commission of Shanghai Municipality (19XD1423000,22ZR1444600); the NYU Shanghai Boost Fund; the China Foreign Experts Program (G2021013002L); the NYU Shanghai Major-Grants Seed Fund; Tamkeen under the NYU Abu Dhabi Research Institute grant CG008. J.S. is supported by the National Natural Science Foundation of China Grant No. 11925507.
\end{acknowledgments}
\setcounter{equation}{0}
\setcounter{figure}{0}
\setcounter{table}{0}
\setcounter{section}{0}
\setcounter{subsection}{0}

\makeatletter
\renewcommand{\theequation}{S\arabic{equation}}
\renewcommand{\thefigure}{S\arabic{figure}}

\begin{widetext}

   \begin{center}
\textbf{\large Supplemental Material}
	\end{center}

	In this Supplemental Material, we briefly explain the Gaussian random process to generate the random realization of our quantum circuits as well as the cost for a hybrid implementation of our scheme.  We also provide details related to the layout of the network architectures that we applied. 
\end{widetext}

\maketitle

\section{ Gaussian random process to generate random circuits}
There are different methods to generate  Gaussian random functions \cite{liu2019advances}. We will explain in detail the one we use. We define a vector $\boldsymbol{\theta}=(\theta(0),\theta(1),\theta (2),...,\theta (2))^T$ and build up the correlation matrix $C$ with elements $C_{nm}=\langle\theta_n \theta_m\rangle=c_0\exp[-(n-m)^2/2\sigma^2)]$, where we assumed a Gaussian correlation function with a correlation length $\sigma$ (though other functional forms could be used). Being real and symmetric, $C$ can be diagonalized as $C=Q\Lambda Q^T$, where $\Lambda$ is a diagonal matrix containing the eigenvalues and $Q$ is an orthogonal matrix. Hence, we can generate the random parameter trajectory as $\boldsymbol{\theta}=Q\sqrt{\Lambda}\boldsymbol{x}$, where the components of $\boldsymbol{x}$ are independent random variables drawn from the unit-width normal distribution ($\langle x_n\rangle=0$ and $\langle x_n x_m\rangle=\delta_{nm}$), which can be easily generated. 

Note that we use qiskit \cite{Qiskit}
for simulating the dynamics of physical observables for random realizations of our circuits.
\section{Hybrid Implementation} 
Here we briefly comment on  the resources required to train our neural network on the data generated by actual experiments.  To calculate the time evolution of any observables at each circuit depth $p$, the experiment needs to be repeated $n$ times for each realization of our random circuit for obtaining an error of $\sim 1/\sqrt{n}$. Hence $n P N_s$ runs are required where $N_s$ is the number of training samples and $P$ is the circuit depth.  Assuming $N_s\sim 5 \times 10^4$, $P\sim 50$ and $n\sim 10^4$ (for a 1 percent projection noise error) on the order of $25 \times 10^{9}$ runs are required.  For a superconducting qubit platform where a single run takes on the order of only a few microseconds, the total run will be on the order of a couple of hours. Note that the number of runs can be still reduced for example by using efficient learning strategies relevant to training the neural networks on noisy measurement data or pre-training the network on simulated data. 

\section{Neural networks layout}
In this section, we present the layout of the architectures that we applied for the dynamics prediction task. We have specified and trained all these different architectures with Keras \cite{chollet2015keras}, a deep-learning framework written for Python. 
\subsection{LSTM neural network}
 In Table. \ref{table_1}, we summarize the details related to the layout of our LSTM  network.  \btxt{This architecture is utilized for the homogeneous (in space) version of both circuits I and II, where $\theta^i_p = \theta^j_p$}.  The training set size for most of the cases that we explored is  60,000. For the last layer, the activation function is ``linear''. As an optimizer, we always use ``adam''.

\begin{table} [!h]
\begin{tabular}{|l|l|l|l|l|}
\hline Layers&&  \#  Neurons  &Activation function\\
\hline Input &LSTM& 2  &- \\
\hline Hidden &LSTM&200  &- \\
\hline Hidden &LSTM& 200  &-\\
\hline Hidden &LSTM& 200  & -\\
\hline Ouput&Dense&  $9[\frac{N-1}{2}]+3$ & Linear \\

\hline
\end{tabular}
\caption{The layout of LSTM neural network for homogeneous circuits where $\theta^i_p = \theta^j_p$. \btxt{ In this case the input to the neural network has a shape of \((\text{number of samples}, 2)\). Each input sample consists of the parameter \(\theta_p\) for each circuit depth \(p\) and the corresponding value of \(p\). The output of the neural network has a shape of \((\text{number of samples}, \text{number of observables})\). The number of observables includes 3 first-order moments and \(9\left\lfloor \frac{N-1}{2} \right\rfloor\) second-order moments, where \(\left\lfloor \cdot \right\rfloor\) denotes the floor function, rounding down to the nearest integer.
}
}
\label{table_1}
\end{table}

\subsection{CONVLSTM neural network}

  In Table. \ref{table_2}, we present the layout of our 1D-CONVLSTM network. Note that we apply the CONVLSTM for the inhomogeneous scenarios  ($\theta^i_p \neq \theta^j_p$) for the circuit I.
\begin{table} [!ht]
\begin{tabular}{|l|l|l|l|l|}
\hline Layers& Filters & Kernel size \\
\hline CONVLSTM1D&70 & 3  \\
\hline CONVLSTM1D& 100& 3  \\
\hline CONVLSTM1D& 100 & 3 \\
\hline CONVLSTM1D& 70 & 3  \\
\hline CONVLSTM1D&  \#  observables & 3  \\
\hline  \multicolumn{3}{|c|}{TimeDistributed(Global max pooling)} \\
\hline
\end{tabular}
\caption{The layout of the 1D-CONVLSTM. CONVLSTM layers capture the temporal-spatial dependencies of the input.  TimeDistributed is a wrapper that applies a layer to every temporal slice of an input.  We use this wrapper with the global max pooling to transfer the input with the temporal-spatial structure to the output with the temporal structure. \btxt{ The input of our 1D-CONVLSTM has a shape defined by (number of samples, number of circuit modules, number of qubits, and number of features). The feature set comprises two elements: the parameter $\theta_p$ for each circuit depth $p$ and the corresponding value of $p$. The output of the neural network has a shape defined by (number of samples, number of circuit modules, number of observables).  The number of qubits and number of modules are specified to "None" denoting variable spatial and temporal dimension enabling extrapolation in size and circuit depth. In this case, in which the circuit is inhomogeneous ($\theta^i_p \neq \theta^j_p$), the network is trained on a selected subset of first and second-order moments of observables. This subset encompasses $3\lfloor N/2\rfloor$ first-order moments and $9\lfloor(N-1)/2\rfloor$) second-order moments. }}
\label{table_2}
\end{table} 

\bibliography{paper}

\end{document}